\documentclass[onecolumn,secnumarabic,amssymb,nobibnotes,aps,prd]{revtex4-1}

\usepackage{xcolor}

\usepackage{graphicx}
\usepackage{booktabs}
\usepackage{amssymb,bm,mathrsfs,bbm,amscd}
\usepackage[tbtags]{amsmath}

\begin{document}

\title{Relativistic formulation for dual one-way Doppler Cancellation Scheme observables for gravitational redshift tests}

\author{Cheng-Gang Qin$^{1}$}
\author{Tong Liu$^{2}$} \email[E-mail:]{liutong2021@csu.ac.cn}
\author{Qin Li$^3$}
\author{Yang Li$^{2}$}
\author{Qi-Long Gong$^1$}
\author{Zhi-Yu Ma$^{3}$}
\author{Yu-Jie Tan$^{3}$}
\author{Cheng-Gang Shao$^{3}$}

\affiliation
{$^{1}$MOE Key Laboratory of TianQin Mission, TianQin Research Center for Gravitational Physics $\&$ School of Physics and Astronomy, Frontiers Science Center for TianQin, Gravitational Wave Research Center of CNSA, Sun Yat-sen University (Zhuhai Campus), Zhuhai 519082, China\\
$^{2}$Key Laboratory of Space Utilization, Technology and Engineering Center for space Utilization, Chinese Academy of Sciences, Beijing 100094, China\\
$^{3}$National Gravitation Laboratory, MOE Key Laboratory of Fundamental Physical Quantities Measurement, and School of Physics, Huazhong University of Science and Technology, Wuhan 430074, People's Republic of China\\}
\date{\today}

\begin{abstract}
Gravitational redshift is a fundamental prediction of general relativity and a sensitive probe of possible deviations from it. Motivated by recent progress in optical clocks and satellite-ground frequency transfer, we study a dual one-way optical Doppler-cancellation observable for space-based gravitational-redshift tests. The corresponding observable is constructed by combining oppositely directed one-way frequency observables, namely an uplink and a downlink whose measurements are recorded at the receiving terminals and combined in post-processing. We derive the corresponding relativistic observable up to order $c^{-3}$, as required for future $10^{-18}$-level clock comparisons. The resulting combination suppresses the first-order Doppler contribution to second order while retaining the gravitational-redshift signal. We analyze the dominant residual effects, including higher-order Doppler terms, atmospheric delay, Shpiro delay, tidal effects, clock synchronization etc. The formalism is applied to representative satellite-ground configurations, including an ACES/CSS-like low-Earth-orbit configuration and a geostationary Earth-orbit satellite. The results show that the dual one-way optical DCS configuration provides a useful relativistic framework and error-modeling reference for future $10^{-18}$-level space-based gravitational-redshift tests, while highlighting the need for stringent clock synchronization.
\end{abstract}

\maketitle

\section{introduction}

General relativity (GR) is the standard theory of gravitation and has successfully passed a broad range of experimental and observational tests \cite{will2014confrontation}. Despite its remarkable success, GR still faces both theoretical and observational challenges. On the theoretical side, a consistent unification of gravity with the other fundamental interactions remains elusive \cite{burgess2004quantum,hossenfelder2013minimal,langacker1981grand}. On the observational side, the observational evidences accumulated over the past few decades, including intriguing puzzles of the dark matter and dark energy, pose potential challenges to GR \cite{RevModPhys.90.045002,RevModPhys.75.559}. High-precision tests of GR therefore remain essential for probing the foundations of gravitational physics and for constraining possible deviations from the standard physics. Clock-comparison experiments are particularly sensitive probes of such possible deviations since they can directly test the relation between proper time and the gravitational field.

The Einstein Equivalence Principle provides a foundation for metric theories of gravity and contains three postulates: the Weak Equivalence Principle (WEP or universality of free fall), Local Lorentz Invariance (LLI), and Local Position Invariance (LPI) \cite{will2014confrontation}. WEP states that in a gravitational field, the trajectory of a freely falling test body is independent of its internal structure and composition. LLI states that the outcome of any local non-gravitational test experiment is independent of the velocity and orientation of the freely-falling reference frame. LPI states that the outcome of any local non-gravitational test experiment is independent of where and when in the universe it is performed. LPI is the topic of the present study. Gravitational redshift is a direct consequence of Local Position Invariance. It states that in the weak-field limit, a clock located at a higher altitude in the Earth's field runs faster than one at a lower altitude. 
Experimental tests of gravitational redshift began with a series of landmark Pound-Rebka-Snider experiments in the 1960s, achieving a 1-10$\%$ precision \cite{PhysRevLett.3.439,PhysRevLett.4.337,PhysRev.140.B788}. Gravity Probe A (GPA) rocket experiment subsequently reached an uncertainty of $1.4\times10^{-4}$ \cite{PhysRevLett.45.2081,vessot1979test}. More recently, analyses of eccentric Galileo satellites obtained uncertainty of $2.48\times10^{-5}$ \cite{PhysRevLett.121.231101,PhysRevLett.121.231102}, and the final RadioAstron analysis confirmed the general-relativistic gravitational-redshift tests with an uncertainty of approximately $4\times10^{-5}$ \cite{nunes2023gravitational,pub.1182157591}. Another experiment using a compact hydrogen maser in a lunar distant retrograde orbit reported a gravitational-redshift test at the $10^{-3}$ level \cite{Li2026gra}. Ground-based experiment of Tokyo Skytree obtained uncertainty of $9.1\times10^{-5}$ \cite{takamoto2020test}. Together, these experiments demonstrate the complementarity of ground-based clock comparisons, Earth-orbiting missions, and experiments extending toward cislunar space.

Rapid advances in optical clocks and optical time-frequency transfer create the potential of improving space-based gravitational-redshift tests substantially \cite{giorgetta2013optical,caldwell2024application,lu2025ntsc,zhang2026liquid,shen2022free,caldwell2025high,zbpb-6qxb,jia2026improved,ren2020development,deng2024cold}. Optical clocks have reached fractional-frequency uncertainties and instabilities near or below $10^{-18}$ in laboratory environments, while transportable clocks and free-space optical links are progressing toward space applications. Proposed or developed space-clock experiments include Atomic Clock Ensemble in Space (ACES) \cite{meynadier2018atomic,savalle2019gravitational}, Space Optical Clock (SOC) \cite{bongs2015development}, China Space Station (CSS) \cite{sun2021test,PhysRevD.108.064031}, China's Lunar exploration project CLEP \cite{qin2024preliminary}, NASA Discovery-class mission VERITAS \cite{PhysRevD.107.064032}, and fundamental physics with a state-of-the-art optical clock in space (FOCOS) \cite{derevianko2022fundamental}. In addition, there are also many attractive studies on spacebased and labbased clocks for redshift \cite{buoninfante2020testing,PhysRevA.110.053102,litvinov2024prospects,PhysRevApplied.21.L061001,ruby2026sensitivity,zheng2023lab,PhysRevD.108.084063,vkyjtxz3,yu2025resolving,belonenko2026precision,tino2019sage}.

The space-based clock-comparison experiments have highly promising potential for improving tests of gravitational redshift, because they can exploit large difference in gravitational potential. A major challenge in such experiments is the first-order Doppler effect, whose fractional magnitude is typically at the level of $10^{-6}$ to $10^{-5}$, much larger than the gravitational redshift signal to be tested. To suppress this contribution, the gravitational-redshift experiments have traditionally relied on the triple-link Doppler cancellation scheme (DCS). 
This scheme generally combined a one-way downlink, referenced to the ground clock, with a coherent two-way link formed by an uplink and a retransmitted downlink. The one-way observable contains both the gravitational redshift signal and the first-order Doppler shift, whereas the coherent two-way observable contains approximately twice the first-order Doppler contribution. Consequently, the first-order Doppler contribution can be reduced to second order by subtracting one-half the two-way observable from the one-way observable, while the gravitational redshift signal is retained. Combining the satellite's orbit determination and speed measurement, a highly accurate test of gravitational redshift was conducted. This strategy was successfully employed in some experiments such as GPA experiments \cite{PhysRevLett.45.2081}. The relativistic descriptions of three-link DCS configurations have been developed to the order of $c^{-3}$ \cite{blanchet2001relativistic}.

Motivated by the development of space-ground optical links and high-performance optical clocks, we provide a relativistic formulation of the dual one-way Doppler cancellation scheme observable for gravitational-redshift tests up to order $c^{-3}$. The term dual one-way denotes two oppositely directed and independently measured one-way links, a one-way uplink and a one-way downlink. The uplink is received and measured at the onboard clock, whereas the downlink is received and measured at the ground clock. Then, the two data are subsequently combined in post-processing. We derive a complete relativistic expression for this dual one-way DCS observable, together with an associated error model suitable for future space missions employing optical clocks at the $10^{-18}$ level. This is particularly relevant because dual-link or oppositely directed frequency-transfer configurations have a clear experimental basis and are relevant to future space-clock missions, while the corresponding $c^{-3}$ relativistic formulation for a dual one-way optical observable, including the effect of clock synchronization between the two receiving terminals, has not been fully established.

Within this framework, we explicitly investigate the contributions from the gravitational redshift, residual Doppler terms, atmospheric effects, Shapiro gravitational effects, tidal effects, and clock-synchronization errors. We then apply the formalism to representative satellite-ground configurations, including the China Space Station (or Atomic Clock Ensemble in Space) and a geostationary Earth-orbit satellite, to quantify the gravitational redshift signal and the dominant residual systematic effects. Compared with conventional three-link DCS formulations, the dual one-way optical DCS configuration provides a compact two-observable framework. This simplification comes with a clock synchronization requirements. In the dual one-way configuration, the uplink and downlink observables are received and time-tagged by different clocks. Their post-processed combination therefore requires knowledge of the relation between the satellite and ground time tags and a common coordinate-time scale, which corresponds to a high-precision synchronization between the satellite and ground clocks. Although optical carriers strongly suppress the ionospheric contribution compared with microwave links, atmospheric effects and link-delay errors must still be modeled or calibrated at the level required by the target redshift sensitivity. In addition, the optical links are assumed to operate during valid observation intervals rather than continuously. Although recent demonstrations of optical time-frequency transfer and clock synchronization support the feasibility of the underlying
concept \cite{chen2024dual,caldwell2023quantum,deschenes2016synchronization,bergeron2019femtosecond,delva2012time,PhysRevA.99.023844,PhysRevApplied.19.054018}, a complete satellite-ground optical-clock experiment at the $10^{-18}$ level has not yet been demonstrated as an integrated space system. The numerical simulations here mainly represent the potential of future satellite-ground clock comparisons for testing gravitational redshift. Improved redshift sensitivity may also provide constraints on physics beyond GR, such as long-range fifth forces, dark matter, or Lorentz violation \cite{PhysRevD.106.095031,PhysRevD.111.064012,roberts2017search,PhysRevD.111.055008,PhysRevD.110.103523}.

The remainder of this paper is organized as follows: Section \ref{sec2} introduce the gravitational redshift and dual one-way Doppler Cancellation Scheme. Section \ref{sec3} introduces the measurement model of dual one-way DCS and discusses the various effects, such as Doppler effect, gravitational redshift, atmospheric frequency shift, clock synchronization etc. Section \ref{sec4} applied dual one-way DCS to analyse the test of gravitational redshift in some missions. Section \ref{sec5} gives the conclusion.

\section{Gravitational Redshift and the Dual One-way Doppler-cancellation Scheme}\label{sec2}

According to Einstein's general relativity, a gravitational field can alter the rate of clocks. In the Earth's gravitational field, a clock at a higher altitude runs faster than one at a lower altitude, giving rise to the gravitational redshift. A commonly used phenomenological parametrization of possible deviations from the GR prediction can be written as
\begin{equation}\label{f1}
  \frac{\Delta f}{f}=(1+\alpha)\frac{\Delta U}{c^2}
\end{equation}
where $\Delta U$ is the difference of gravitational potential, $c$ is the vacuum speed of light, and $\alpha$ parametrizes a possible violation of the GR prediction, with $\alpha=0$ in GR. This effect can be tested by comparing the frequencies or times of two clocks with electromagnetic signal exchange. A larger gravitational-potential difference generally leads to a stronger redshift signature and therefore improves the sensitivity of the test. For this reason, satellite-ground clock comparisons provide a particularly favorable configuration for high-precision gravitational-redshift experiments, especially when combined with state-of-the-art optical atomic clocks.

\begin{figure}
\includegraphics[width=0.7\textwidth]{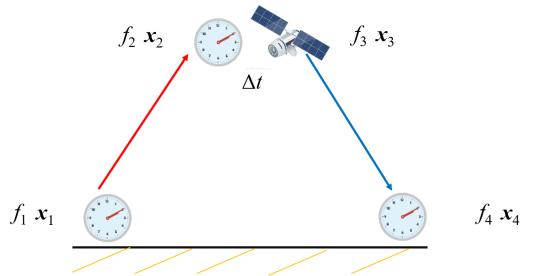}
\caption{\label{fig1} Diagram of dual one-way Doppler cancellation scheme. The red line represents the uplink signal and blue line indicates the downlink signal. The term ``dual one-way" denotes two oppositely directed and independently measured one-way links, not a single unidirectional link. The uplink is received and measured at the onboard clock, whereas the downlink is received and measured at the ground clock. Then, the two data are subsequently combined in post-processing.}
\end{figure}

We consider here a dual one-way Doppler cancellation scheme(dual one-way DCS), as illustrated in Fig. \ref{fig1}. The term ``dual one-way" denotes two oppositely directed and independently measured one-way links: an uplink (red line) from the ground clock $A$ to the satellite clock $B$, and a downlink (blue line) from the satellite clock $B$ to the ground clock $A$. The uplink is received and measured at the onboard clock, whereas the downlink is received and measured at the ground clock. Two one-way links yield two sets of data. Then, the two data are subsequently combined in post-processing.

For the uplink, at coordinate time $t_1$, the ground clock $A$ at position $\boldsymbol{{x}}_{1}$ emits an optical signal with frequency $f_1$. This signal propagates freely through space and is received at coordinate time $t_2$ by satellite clock $B$ at position $\boldsymbol{{x}}_{2}$. The satellite clock $B$ measures the received frequency as $f_2$, from which the uplink observable $y_{\text{up}}$ is constructed.

For the downlink, at coordinate time $t_3$, the satellite clock $B$ at position $\boldsymbol{{x}}_{3}$ sends a light signal towards the ground station with frequency $f_3$. After free-space propagation, ground clock $A$ at position $\boldsymbol{{x}}_{4}$ receives this signal at coordinate time $t_4$. The ground clock $A$ measures the received frequency as $f_4$. The corresponding downlink observable is denoted by $y_{\text{do}}$.

In general, the two one-way measurements are not exactly synchronous. We therefore introduce the time offset
\begin{equation}\label{001}
\Delta t=t_3-t_2,
\end{equation}
which characterizes the mismatch between the uplink reception event and the downlink emission event. In the ideal synchronous limit, $\Delta t=0$. In realistic experiments, however, a small but finite $\Delta t$ remains and must be taken into account, since it leads to a correction in the combined observable.

The dual one-way scheme thus produces two independent frequency-comparison observables, one from the uplink and the other from the downlink. The uplink observable  $y_{\text{up}}$ contains both the first-order Doppler and gravitational redshift. The downlink observable  $y_{\text{up}}$ includes the first-order Doppler and ``negative" gravitational redshift (or ``gravitational blueshift"). By constructing an appropriate linear combination of $y_{\text{up}}$ and $y_{\text{do}}$, one can obtain the measurement of dual one-way DCS as follows
\begin{eqnarray}\label{002}
y_{\text{dcs}}&=&y_{\text{do}}-y_{\text{up}},
\end{eqnarray}
in which the first-order Doppler contribution can be strongly suppressed, while the gravitational redshift is enhanced by a factor of two.

\section{Observable of the Dual One-way Doppler Cancellation Scheme}\label{sec3}

We consider a clock-comparison experiment between a ground clock and a satellite clock in the Geocentric Celestial Reference System (GCRS). To construct the observable of the dual one-way DCS, it is needed to derive the one-way frequency-comparison observables for uplink and downlink. We take the uplink as an example for calculating the clock comparison, and the calculation for downlink is similar. Generally, the difference in frequencies between the two clocks can be characterized by the frequency shift $f_2/f_1$. Let a wavefront of the signal be emitted from the position $\boldsymbol{{x}}_{1}$ at coordinate time $t_1$ and proper time $\tau_1$, and received at the position $\boldsymbol{{x}}_{2}$ at coordinate time $t_2$ and proper time $\tau_2$. The emission and reception events are related by the light-travel-time equation, so that $t_1$ is an implicit function of $t_2$. Denoting the electromagnetic phase by $\phi$, the proper frequencies measured at emission and reception are denoted by $f_1=(2\pi)^{-1}d\phi/d\tau_1$ and $f_2=(2\pi)^{-1}d\phi/d\tau_2$, respectively. Since the same phase is transferred along the ray, their ratio, or the one-way frequency shift of the uplink, can be written as
\begin{equation}\label{gr1}
\frac{f_2}{f_{1}}=\frac{d \tau_1}{dt_1}\left(\frac{d \tau_2}{d t_2}\right)^{-1} \frac{dt_1}{dt_2}.
\end{equation}
This equation is composed of a product of three fractional factors. The first two factors describe the relationship between coordinate time and the proper times of the emitter and receiver, and contain gravitational redshift, time dilation and tidal effects. The last factor is the derivative of the emission time with respect to the reception time and describes the propagation contribution, including the Doppler effect, atmospheric effect, Shapiro-delay contributions etc.

Firstly, we calculate the clock-dependent part (first two factors) in the GCRS, which includes gravitational redshift that we are concerned about. This part can be calculated using the invariance of Riemannian space-time intervals. Considering the metric tensor of the GCRS, it is given by
\begin{eqnarray}\label{grcd1}
\frac{d \tau_1}{dt_1}\left(\frac{d \tau_2}{d t_2}\right)^{-1}=1-\frac{1}{c^2}\left(U_{E}(\boldsymbol{{x}}_{1})-U_{E}(\boldsymbol{{x}}_{2})
+u_{\text{tid}}(\boldsymbol{{x}}_{1})-u_{\text{tid}}(\boldsymbol{{x}}_{2})
+\frac{\boldsymbol{{v}}_{g}^2(t_{1})}{2}-\frac{\boldsymbol{{v}}_{s}^2(t_{2})}{2}\right),
\end{eqnarray}
where $U_{E}$ is the gravitational potential of the Earth, $u_{\text{tid}}$ is the tidal potential whose dominant contributions arise from the Moon and the Sun, $\boldsymbol{{v}}$ is the velocity of the ground station or satellite, $\boldsymbol{{v}}_{g}(t_{1})$ is the velocity of ground station at time $t_{1}$, $\boldsymbol{{v}}_{s}(t_{2})$ is the velocity of satellite at time $t_{2}$.

Then, we consider the path-dependent part (third factor of Eq.(\ref{gr1})) which is related to the transmission path of signal. This part can be calculated by the coordinate times between signal's emission and reception. In the uplink, the signal's emission time and reception time can be written as
\begin{equation}\label{gr2}
t_{1}=t_{2}-T_{12}-\Delta_{12}^{\text{atm}}-\Delta_{12}^{\text{Shap}},
\end{equation}
where $T_{12}=R_{12}/c=|\boldsymbol{{x}}_{2}-\boldsymbol{{x}}_{1}|/c$ is the travel time of the signal in Euclidean space, $\Delta_{12}^{\text{atm}}$ is the atmospheric
delay, and $\Delta_{12}^{\text{Shap}}$ is the Shapiro gravitational delay. From the time interval Eq.(\ref{gr2}) between the transmission and reception, the frequency shift of path-dependent part of uplink becomes
\begin{equation}\label{gr3}
\frac{dt_1}{dt_2}=1-\frac{d T_{12}}{dt_2}-\frac{d \Delta_{12}^{\text{atm}}}{dt_{2}}-\frac{d \Delta_{12}^{\text{Shap}}}{dt_2},
\end{equation}
where the second term represents the Dopper contribution caused by the relative motion between the ground station and the satellite, the third term is the atmospheric frequency shift, and the fourth term is the gravitational frequency shift caused by Shapiro delay.

Combining Eqs.(\ref{gr1}) to (\ref{gr3}), we can express the frequency-comparison observable (or one-way frequency transfer) of the the uplink as
\begin{eqnarray}\label{gr4}
y_{\text{up}}=\frac{f_2}{f_1}&=&1-\frac{d R_{12}}{cdt_2}-\frac{d \Delta_{12}^{\text{atm}}}{dt_{2}}-\frac{d \Delta_{12}^{\text{Shap}}}{dt_2}+\frac{1}{c^2}\Big{(}U_{E}(\boldsymbol{{x}}_{2})-U_{E}(\boldsymbol{{x}}_{1})\nonumber\\
&+&u_{\text{tid}}(\boldsymbol{{x}}_{2})-u_{\text{tid}}(\boldsymbol{{x}}_{1})
+\frac{\boldsymbol{{v}}_{s}^2(t_{2})}{2}-\frac{\boldsymbol{{v}}_{g}^2(t_{1})}{2}\Big{)},
\end{eqnarray}
For typical satellite-ground clock comparisons, the velocity of satellites is about $3\sim8$ km/s, and the first-order Doppler shift is the order of $10^{-5}$. For evaluating the frequency comparison to the level of $10^{-18}$,  the corresponding requirement on velocity measurement is about or better than $10^{-9}$ m/s, which is beyond current technological capability. So, a Doppler cancellation scheme is applied to suppress Doppler shift to the second order. By comparison, the atmospheric and Shapiro-induced frequency shifts can each reach the $10^{-14}$ level. The gravitational redshift term depends on the satellite's orbit and the location of the ground station. For the ACES and CSS missions, the gravitational redshift is about $4\times10^{-11}$. For the satellite-ground clock comparisons at higher orbits, the gravitational redshift can reach several $10^{-10}$. With the clock comparison at the level of $10^{-18}$, the gravitational redshift tests could in principle achieve sensitivities ranging from $10^{-6}$ to $10^{-9}$. The two most significant challenges are: achieving $10^{-18}$-level satellite-ground clock comparison in practice, and modeling all relevant effects to the same level.

Following the same procedure, we obtain the frequency-comparison observable of the the downlink
\begin{eqnarray}\label{gr5}
y_{\text{do}}=\frac{f_4}{f_3}&=&1-\frac{d R_{34}}{cdt_3}-\frac{d \Delta_{34}^{\text{atm}}}{dt_{4}}-\frac{d \Delta_{34}^{\text{Shap}}}{dt_4}+\frac{1}{c^2}\Big{(}U_{E}(\boldsymbol{{x}}_{4})-U_{E}(\boldsymbol{{x}}_{3})\nonumber\\
&+&u_{\text{tid}}(\boldsymbol{{x}}_{4})-u_{\text{tid}}(\boldsymbol{{x}}_{3})
+\frac{\boldsymbol{{v}}_{g}^2(t_{4})}{2}-\frac{\boldsymbol{{v}}_{s}^2(t_{3})}{2}\Big{)},
\end{eqnarray}
This equation can be used to estimate various effects of the downlink. Compared with the uplink observable in Eq.(\ref{gr4}), the downlink contains an almost identical first-order Doppler contribution, whereas the gravitational-redshift term appears with the opposite sign. This property enables the construction of the dual one-way DCS observable
\begin{eqnarray}\label{gr6}
y_{\text{dcs}}&=&y_{\text{do}}-y_{\text{up}}\nonumber\\
&=&y^{\text{gr}}_{\text{dcs}}+y^{\text{dop}}_{\text{dcs}}+y^{\text{atm}}_{\text{dcs}}+y^{\text{Shap}}_{\text{dcs}}+y^{\text{tid}}_{\text{dcs}}+
  y^{\text{c}}_{\text{dcs}}+y^{\text{n}}_{\text{dcs}}.
\end{eqnarray}
On the right-handle of Eq.(\ref{gr6}), the first term represents the gravitational redshift contribution. In the dual one-way DCS, this contribution is twice that of a one-way frequency transfer, which is advantageous for improving the precision of gravitational-redshift tests. The second term represents the residual Doppler effect with the $c^{-2}$ order and higher-order terms, where the higher-order terms contain time dilation and coupling terms of the Doppler effect with other effects. Due to the imperfect symmetry between the uplink and downlink paths, the first-order Doppler effect is not completely canceled after Doppler cancellation scheme, but is instead suppressed to a second order. The third and fourth terms represent the residual atmospheric contribution and the Shapiro gravitational delay contribution, respectively. Similar to the Doppler term, these effects are significantly reduced compared with those in one-way frequency transfer. The fifth term represents the tidal effect. The sixth term represents the influences of clock synchronization and onboard time delay. The last term is the clock frequency noise.

In the following, we analyze and evaluate each term of these effects separately.

\subsection{Gravitational redshift}

The gravitational redshift is the primary effect targeted in the satellite-ground clock comparisons for testing general relativity's gravitational redshift prediction. In such experiments, the dominant contribution arises from the Earth's gravitational field. High-precision clock comparison therefore requires an accurate modeling of the terrestrial gravitational potential. The current Earth gravitational field model is given by Earth Gravitational Model 2008 (EGM2008) \cite{pavlis2012development}. The Earth's gravitational potential is expanded in terms of spherical harmonic coefficients as follows
\begin{equation}\label{grgr1}
  U_E(\boldsymbol{{x}})
  =\frac{GM_E}{r}\sum_{l=0}^\infty\sum_{k=0}^l\left(\frac{r_{0E}}{r}\right)^l P_{lk}\left(\cos\theta\right)
  \left[C_{lk}\cos\left(k\varphi\right)+S_{lk}\sin\left(k\varphi\right)\right],
\end{equation}
where $GM_E$ is the Earth's gravitational constant, $P_{lk}(\cos\theta)$ are the associated Legendre functions, $r_{0E}$ is the Earth's equatorial radius, $\theta$ is latitude, $\varphi$ is longitude, $C_{lk}$ and $S_{lk}$ are the spherical harmonic coefficients determined through some gravity data. For convenience, we represent the gravitational potential $ U_E(\boldsymbol{{x}})$ of EGM2008 in the for $GM_{E}f(r,\theta,\varphi)/r$. The uncertainty of EGM2008 in calculating the gravitational potential is typically at the level of about $1$ m$^2$/s$^2$, which corresponds to a fractional frequency uncertainty of approximately $1\times10^{-17}$. By combining the gravity data or more precise measurements, this uncertainty may be reduced to the level of  $0.1$ m$^2$/s$^2$.

By combining the EGM2008 model and one-way frequency transfer (\ref{gr4}), the gravitational redshift for the uplink can be derived as
\begin{equation}\label{grgr2}
   y^{\text{gr}}_{\text{up}}
   =\frac{GM_E}{c^2r_2}f(r_2,\theta_2,\varphi_2)-\frac{GM_E}{c^2r_1}f(r_1,\theta_1,\varphi_1)
\end{equation}
where $f(r_1,\theta_1,\varphi_1)$ and $f(r_2,\theta_2,\varphi_2)$ represent the EGM2008's expansions for positions $\boldsymbol{{x}}_1$ and $\boldsymbol{{x}}_2$, $r_{1}=|\boldsymbol{{x}}_{1}|$ and $r_{2}=|\boldsymbol{{x}}_{2}|$ are the positions of the ground station at coordinate time $t_1$ and satellite at coordinate time $t_2$, respectively. For most satellite-ground clock-comparison configurations, the gravitational redshift contribution is typically of order $10^{-11}$ to $10^{-10}$. For example, the gravitational redshift is about $-4\times10^{-11}$ for ACES or CSS missions, and about $-5\times10^{-10}$ for GPS-like orbit. With the state-of-the-art optical clocks, this would allow gravitational-redshift tests at the $10^{-8}$ level.

Similarly, the gravitational redshift in the downlink $y^{\text{gr}}_{\text{do}}$ is given by $GM_E f(r_4,\theta_4,\varphi_4)/(c^2 r_4)-GM_E f(r_3,\theta_3,\varphi_3)/(c^2 r_3)$, which is approximately equal to $-y^{\text{gr}}_{\text{up}}$ in the near-synchronous case. With this characteristic, the gravitational redshift contribution in the dual one-way DCS observable is by
\begin{equation}\label{grgr3}
   y^{\text{gr}}_{\text{dcs}}=y^{\text{gr}}_{\text{do}}-y^{\text{gr}}_{\text{up}}
   \simeq 2\frac{GM_E}{c^2r_4}f(r_4,\theta_4,\varphi_4)-2\frac{GM_E}{c^2r_3}f(r_3,\theta_3,\varphi_3),
\end{equation}
where the approximation holds when the uplink and downlink are nearly synchronized. Therefore, in the dual one-way DCS, the gravitational redshift signal is doubled relative to that in a single one-way link. It implies that this scheme is sensitive to gravitational redshift.

\subsection{Doppler effect}
The Doppler effect is the dominant contribution in satellite-ground clock comparisons, arising from the relative velocity between the satellite and the ground station. Its magnitude is much larger than that of the gravitational redshift. Therefore, a Doppler-cancellation scheme is essential for high-precision clock comparison. To illustrate the origin of this effect and the degree to which it can be suppressed, we first consider the Doppler effect on the uplink.

According to the Appendix.\ref{app1}, the Doppler contribution to the uplink can be written as
\begin{eqnarray}\label{grd1}
y^{\text{dop}}_{\text{up}}&=&-\frac{\boldsymbol{{N}}_{12}\cdot\boldsymbol{{v}}_{12}}{c}
-\frac{(\boldsymbol{{N}}_{12}\cdot\boldsymbol{{v}}_{12})(\boldsymbol{{N}}_{12}\cdot\boldsymbol{{v}}_{g}(t_1))}{c^2}+\frac{1}{c^{2}}\left(\frac{\boldsymbol{v}_{s}^2(t_2)}{2} - \frac{\boldsymbol{v}_{g}^2(t_1)}{2}  \right) \nonumber\\
&-&\frac{\boldsymbol{{N}}_{12}\cdot\boldsymbol{{v}}_{12}}{c^3}\left[(\boldsymbol{{N}}_{12}\cdot\boldsymbol{{v}}_{g}(t_1))^2
+\frac{\boldsymbol{{v}}^2_{s}(t_2)}{2}-\frac{\boldsymbol{{v}}^2_{g}(t_1)}{2}+U_E(\boldsymbol{{x}}_{s}(t_2)) -U_E(\boldsymbol{{x}}_{g}(t_2))\right],
\end{eqnarray}
where $\boldsymbol{{N}}_{12}=\boldsymbol{{R}}_{12}/R_{12}$ is the unit vector of coordinate distance $\boldsymbol{{R}}_{12}$, and $\boldsymbol{{v}}_{12}=\boldsymbol{{v}}_{s}(t_2)-\boldsymbol{{v}}_{g}(t_1)$ is the relative velocity between satellite and ground station. The first term is the first-order Doppler shift, while the remaining terms represent higher-order terms, which includes higher-order Doppler, time dilation and coupling terms of the Doppler effect with other effects. Typically, the relative velocity between the satellite and ground station ranges from 3$\sim$8 km/s, corresponding to a Doppler shift with the order of $10^{-5}$. This is several orders of magnitude larger than the gravitational-redshift signal and thus must be suppressed in precision clock comparison. For the downlink, the corresponding Doppler frequency shift can be derived through an analogous process, achieved by substituting $1,2$ and $s,g$ in Eq. (\ref{grd1}) with $3,4$ and $g,s$, respectively.

By combining the uplink and downlink observables in the dual one-way DCS, the first-order Doppler terms are strongly canceled. This cancellation originates from the fact that the first-order Doppler shift is determined by the relative velocity between the satellite and the ground station. For two nearly symmetric one-way links, the uplink and downlink probe almost the same relative motion. Consequently, the first-order Doppler terms of the uplink and downlink enter the DCS combination and cancel to a higher order. The residual Doppler observable is therefore no longer of order $c^{-1}$, but is instead dominated by second-order terms  $c^{-2}$. This is the essential reason why the dual one-way DCS is effective for precision gravitational-redshift tests.

In practice, however, the measured data general are time-tagged at the reception instants $t_2$ and $t_4$. It is therefore convenient, and in fact necessary for data analysis, to re-express all quantities in terms of variables referenced to the reception time. In actual measurements, the recorded observables are naturally associated with reception time tags rather than emission time tags. If the Doppler model were written in terms of quantities evaluated at different coordinate times along the light path, an additional mismatch would be introduced in the data processing. Although such a mismatch is negligible at moderate precision, it becomes significant for clock comparison at the $10^{-18}$ level. Expressing all quantities in terms of reception time therefore ensures both theoretical consistency and direct compatibility with the experimentally available data.

To this end, we introduced some quantities at the reception time such as the instantaneous distance $\boldsymbol{{D}}_{34}=\boldsymbol{{x}}_{g}(t_{4})-\boldsymbol{{x}}_{s}(t_{4})$, satellite velocity $\boldsymbol{{v}}_{s}(t_{4})$, ground station velocity $\boldsymbol{{v}}_{g}(t_{4})$, and ground station acceleration $\boldsymbol{{a}}_{g}(t_{4})$ at coordinate time $t_4$.
According to the Appendix.\ref{app1}, after rewriting the uplink contribution in terms of quantities evaluated at the common reception time and combining it with the downlink, the residual Doppler shift in the dual one-way DCS can be written as
\begin{eqnarray}\label{grd2}
y^{\text{dop}}_{\text{dcs}}=\frac{(\boldsymbol{{n}}_{34}\cdot\boldsymbol{{v}}_{34}(t_4))^2}{c^2}-\frac{2\boldsymbol{{D}}_{34}\cdot\boldsymbol{{a}}_{g}(t_4)}{c^2}
-\frac{\boldsymbol{{v}}^{2}_{34}(t_4)}{c^2}
+y^{\text{dop3}}_{\text{dcs}},
\end{eqnarray}
where $\boldsymbol{{n}}_{34}=\boldsymbol{{D}}_{34}/|\boldsymbol{{D}}_{34}|$ is the unit vector of the instantaneous distance $\boldsymbol{{D}}_{34}$, $\boldsymbol{{v}}_{34}(t_4)=\boldsymbol{{v}}_{g}(t_4)-\boldsymbol{{v}}_{s}(t_4)$ represents the instantaneously relative velocity of satellite and ground station at coordinate time $t_4$. Note that all quantities in Eq.(\ref{grd2}) are expressed as the values at time $t_{4}$. From Eq.(\ref{grd2}), the residual Doppler shift is seen to be of order $10^{-10}$, much smaller than the first-order Doppler term but still much larger than the targeted $10^{-18}$-level precision if left unmodeled.

It should be emphasized that the instantaneous distance $\boldsymbol{D}_{34}$ and the coordinate distance $\boldsymbol{R}_{34}$ are not identical. Replacing $\boldsymbol{D}_{34}$ with $\boldsymbol{R}_{34}$ in the data analysis would lead to a non-negligible modeling error, which can exceed the $10^{-18}$ level. This point is important in practice, because the coordinate distance $\boldsymbol{R}_{34}$ depends on the emission and reception events $(t_3,t_4)$, whereas the experimentally recorded observables are naturally referenced to the reception time. For example, in the clock comparison of ACES, the difference between $\boldsymbol{{D}}_{34}$ and $\boldsymbol{{R}}_{34}$ can lead to an estimated error of $5\times10^{-18}$ in the residual Doppler effect. Therefore, all quantities entering the Doppler model should be consistently expressed in terms of reception-time variables such as $\boldsymbol{D}_{34}$, $\boldsymbol{n}_{34}$, and $\boldsymbol{v}_{34}(t_4)$.

In addition, the residual third-order Doppler $y^{\text{dop3}}_{\text{dcs}}$ in Eq.(\ref{grd2}) is given by
\begin{equation}\label{redop333}
\begin{aligned}
 y^{\text{dop3}}_{\text{dcs}}& =2\frac{\boldsymbol{D}_{34}\cdot\boldsymbol{b}_{g}(t_4)}{c^3}D_{34}
 +4\frac{\boldsymbol{a}_{g}(t_4)\cdot\boldsymbol{v}_{g}(t_4)D_{34}}{c^3}
 -2\frac{\boldsymbol{a}_{s}\left(t_4\right)\cdot\boldsymbol{v}_{34}(t_4)D_{34}}{c^3}
  +\frac{\left(\boldsymbol{n}_{34}\cdot\boldsymbol{v}_{34}(t_4)\right)}{c^3}\times \\
 & \biggl[3\Big(\boldsymbol{n}_{34}\cdot\boldsymbol{v}_{{g}}(t_4)\Big)^2+3\Big(\boldsymbol{n}_{34}\cdot\boldsymbol{v}_{s}\left(t_4\right)\Big)^2
 -\boldsymbol{v}_{34}^2(t_4)-2\boldsymbol{v}_{s}^2(t_4)+\boldsymbol{a}_{g}(t_4)\cdot\boldsymbol{D}_{34}
 +2\boldsymbol{a}_{s}(t_4)\cdot\boldsymbol{D}_{34}\biggr]  \\
&+ \frac{\boldsymbol{{n}}_{34}\cdot\boldsymbol{{v}}_{34}(t_4)}{c^3}\left[ {\boldsymbol{v}_{s}^2(t_4)}-{\boldsymbol{v}_{g}^2(t_4)}+2U_E(\boldsymbol{{x}}_{s}(t_4)) -2U_E(\boldsymbol{{x}}_{g}(t_4))\right],
\end{aligned}
\end{equation}
where all quantities are also expressed as the values at coordinate time $t_{4}$, $\boldsymbol{b}=d \boldsymbol{a}/dt$ denotes the time derivative of the acceleration, the last term holds under the approximation of near-simultaneity of the upperlink and downlink, and the tidal contribution to the scalar potentials has been neglected. For satellite-ground clock comparison with the level of $10^{-18}$, the contribution of residual third-order Doppler erm cannot be ignored and must be deducted during the data analysis.

\subsection{Atmospheric frequency shift}

In satellite-ground clock comparisons, the signals traversing the Earth's atmosphere experience additional propagation delays. For optical links, the dominant contribution is usually the neutral-atmospheric delay, while the dispersive ionospheric delay is strongly suppressed but should be considered separately. The neutral-atmospheric delay can be expressed as
\begin{equation}\label{gram001}
 \Delta^{\text{atmo}}=\frac{d_{\text{atmo}}}{c}=\frac{1}{c}\int N dl,
\end{equation}
where $N$ is the group refractivity and the integration is calculated along the propagation path. In the Marini-Murray model or the Mendes-Pavlis model \cite{mendes2004high,ciddor1996refractive},
this delay can be typically modeled with a zenith propagation delay $d_{\text{atmo}}^z$ and a mapping function $m(\epsilon)$, thus is written as the form $d_{\text{atmo}}=m(\epsilon)d_{\text{atmo}}^z$. Temporal variations in the neutral-atmospheric delay induce an neutral-atmospheric frequency shift in the clock comparison. Considering the uplink, the neutral-atmospheric frequency shift can be expressed as
\begin{equation}\label{gram1}
y^{\mathrm{atmo}}_{\mathrm{up}}
=
-\frac{d\Delta^{\mathrm{atmo}}_{12}}{dt_2}
=
-\frac{1}{c}\frac{d}{dt_2}\!\left[m(\epsilon_u)d_{\mathrm{atm}}^{z}\right],
\end{equation}
where $\epsilon_u$ is the uplink elevation angle. Using the Mendes-Pavlis model, the zenith delay can be expressed in terms of the station meteorological parameters, yielding the uplink neutral-atmospheric frequency shift
\begin{equation}\label{gram2}
  y^{\text{atmo}}_{\text{up}}
  =-\frac{d}{cdt_2} \left[10^{-6}m(\epsilon_{u}) \left(4.16579\frac{f_{h}(\lambda)}{f(\phi,H)}P_{s}+(5.316f_{nh}(\lambda)-3.759f_{h}(\lambda))\frac{e_{s}}{f(\phi,H)}\right)\right],
\end{equation}
where $\lambda$ is the vacuum wavelength ($\mu$m) of the signal, $\phi$ is the latitude of ground station, $H$ is the height (km) of ground station, $P_{s}$ is the station's surface barometric pressure, $e_{s}$ is the station's surface water vapor pressure. The $f_{h}(\lambda)$, $f(\phi,H)$ and $f_{nh}(\lambda)$ are the integration functions, whose expressions can be found in Ref.\cite{mendes2004high}. For a simple order-of-magnitude estimate, the atmospheric frequency shift of the one-way link may be reach the level of $10^{-10}$. From Eq.(\ref{gram2}), it is clear that since the uplink and downlink traverse nearly the same atmospheric column, their neutral-atmospheric frequency shifts are nearly equal.

Considering dual one-way DCS, the residual neutral-atmospheric frequency shift is given by
\begin{equation}\label{gram3}
  y^{\text{atmo}}_{\text{dcs}}
  =\frac{d}{cdt_4} \left[10^{-6}(m(\epsilon_{d})- m(\epsilon_{u})) \left(4.16579\frac{f_{h}(\lambda)}{f(\phi,H)}P_{s}+(5.316f_{nh}(\lambda)-3.759f_{h}(\lambda))\frac{e_{s}}{f(\phi,H)}\right)\right],
\end{equation}
where $\epsilon_{d}$ is the elevation angle of downlink signal, and since the time interval between $t_2$ and $t_4$ is small, it is sufficient to use the derivative of $t_4$ for the uplink neutral-atmospheric frequency shift. Since the asymmetry between the uplink and downlink is much small, the residual neutral-atmospheric frequency shift may be smaller than $10^{-18}$.

We have so far considered the neutral-atmospheric contribution. The ionospheric contribution should be treated separately. To leading order, the ionospheric delay of a one-way link is
\begin{equation}
\Delta^{\text{iono}} =
\frac{40.3}{c f^2}\,{\text{TEC}_{\text{slant}}},
\end{equation}
where $f$ is the carrier frequency and ${\text{TEC}}_{\text{slant}}$ is the slant total electron content. The corresponding fractional frequency shift in the uplink is
\begin{equation}
y^{\text{iono}}_{\text{up}}=
-\frac{40.3}{c f^2}
\frac{d\,{\text{TEC}_{\text{slant}}}}{dt_2}.
\end{equation}
For the dual one-way DCS, the residual ionospheric term can be written as
\begin{equation}
y^{\text{iono}}_{\text{dcs}}\simeq
-\frac{40.3}{c}
\frac{d}{dt_4}
\left(
\frac{{\text{TEC}^{u}_{\text{slant}}}}{f_u^2}
-
\frac{{\text{TEC}^{d}_{\text{slant}}}}{f_d^2}
\right).
\end{equation}
For optical frequencies, this term is strongly suppressed by the factor $f^{-2}$. Taking $f\sim 10^{14}$ Hz, one TECU corresponds to a delay of order $10^{-9}$ m, or a time delay of order $10^{-18}$ s. Thus, for the optical links considered in this work, the ionospheric frequency shift is smaller than $10^{-18}$ level, far below the dominant neutral-atmospheric term. This is different from microwave links such as ACES-MWL, where an additional frequency link is used mainly to calibrate the ionospheric delay. If the present scheme were implemented with microwave or mixed-frequency links, an additional ionospheric calibration channel or external TEC correction would be required.

\subsection{Shapiro Gravitational frequency shift}

From Einstein's general relativity, the gravitational field can lead to an extra time delay in light travel time, known as Shapiro delay. In the clock comparison, the temporal variations in the Shapiro delay induce a Shapiro gravitational frequency shift. Assuming the metric is given, the standard methods to derive Shapiro delay are solving the null geodesic equation or the eikonal equation \cite{PhysRevD.94.124007,ashby2010accurate}, and different approaches are alternative based on the Synge World function and time transfer function \cite{le2004world,teyssandier2008general,PhysRevD.89.064045,PhysRevD.89.064045}. In the post-Newtonian approximation, the metric is expanded as $g^{\alpha\beta}=\eta^{\alpha\beta}+h^{\alpha\beta}$ with $\eta_{\alpha\beta}=$diag($-$1,+1,+1,+1), and the gravitational perturbation $h_{\alpha\beta}$. The Shapiro delay of light propagation in the gravitational field can be calculated by $\Delta^{\text{Shap}} = ({R}/{2c}) \int_0^1 \left(h^{00}-2n^i h^{0i}+n^i n^j h^{ij}\right)_{x_{(l)}} dl$, where the integral is calculated along the straight line between the ground station and satellite. Considering the uplink signal, the Shapiro delay can be calculated by the uplink path as
\begin{equation}\label{grgt1}
\begin{split}
\Delta ^{\text{Shap}}_{\text{up}} =  \frac{2GM_E}{c^3}\ln\frac{r_1+r_2+R_{12}}{r_1+r_2-R_{12}} .
\end{split}
\end{equation}
From this equation, the Shapiro time delay contributed by the Earth can reach the level of $10^{-10}$ s.
Combining Eqs. (\ref{gr4}) and (\ref{grgt1}), the Shapiro gravitational frequency shift of the uplink is given by
\begin{equation}\label{grgt2}
\begin{split}
y^{\text{Shap}}_{\text{up}} = & - \frac{GM_E(r_{1}+r_{2})}{c^3 r_{1} r_{2}} \bigg{[} \left(\frac{2}{1+\boldsymbol{{N}}_{1}\cdot \boldsymbol{{N}}_{2}}-\frac{r_{1}-r_{2}}{r_{1}+r_{2}} \right) \boldsymbol{{N}}_{12}\cdot(\boldsymbol{{v}}_{g}(t_1)-\boldsymbol{{v}}_{s}(t_2)) \\
&+ \frac{2R_{12}}{r_{1}+r_{2}}\frac{\boldsymbol{{N}}_{1}\cdot \boldsymbol{{v}}_{g}(t_1)+\boldsymbol{{N}}_{2}\cdot \boldsymbol{{v}}_{s}(t_2)}{1+\boldsymbol{{N}}_{1}\cdot \boldsymbol{{N}}_{2}}\bigg{]},
\end{split}
\end{equation}
where $\boldsymbol{{N}}_{1}=\boldsymbol{{x}}_{1}/r_1$ and $\boldsymbol{{N}}_{2}=\boldsymbol{{x}}_{2}/r_2$. For an order-of-magnitude estimate, the gravitational frequency shift of the uplink can reach the level of $10^{-14}$. One can also derives the gravitational frequency shift $y^{\text{Shap}}_{\text{do}}$ for the downlink by substituting 1,2 and $g,s$ in Eq.(\ref{grgt2}) with 3,4 and $s,g$, respectively. The values of gravitational frequency shift are approximately equal for the uplink and downlink signals.

In the measurement of dual one-way DCS, the residual gravitational frequency shift is given by
\begin{equation}\label{grgt3}
  y^{\text{Shap}}_{\text{dcs}}=y^{\text{Shap}}_{\text{do}}-y^{\text{Shap}}_{\text{up}}
\end{equation}
This equation can be used to calculate the influences of Shapiro delay in the clock comparisons. Since the Shapiro time delays of the uplink and downlink are nearly identical, the residual gravitational frequency shift is much smaller than $10^{-14}$.

\subsection{Tidal influences}
Similar to the gravitational redshift, tidal potentials also induce slight variations in clock frequency. The tidal potential $u_{\text{tid}}(\boldsymbol{{x}})$ can be decomposed into two components, the external mass component $u^{\text{ext}}_{\text{tid}}(\boldsymbol{{x}})$ and the internal mass component $u^{\text{int}}_{\text{tid}}(\boldsymbol{{x}})$. The external part is primarily contributed by the tidal potentials of the Sun and Moon, in the conventional Newtonian tidal potential form. The internal part is mainly from Earth's tidal deformations, which comprises solid Earth tides, ocean tides, and pole tides. Solid Earth tides are the main contribution for the internal part, which generates surface displacements of approximately 10$\sim$30 cm/day.
The external and internal tidal potentials can be expressed as \cite{petit2010iers}
\begin{equation}\label{grti1}
  u^{\text{ext}}_{\text{tid}}(\boldsymbol{{x}})=\sum_{b\neq E}\left( U_{b}(\boldsymbol{{r}}_{bE}+\boldsymbol{{x}})
  -U_{b}(\boldsymbol{{r}}_{bE})-\boldsymbol{{x}}\cdot \nabla U_{b}(\boldsymbol{{r}}_{bE})\right)
\end{equation}
and
\begin{equation}\label{grti2}
  u^{\text{int}}_{\text{tid}}(\boldsymbol{{x}})=
  \frac{GM_E}{r}\sum_{l=0}^\infty\sum_{k=0}^l\left(\frac{r_{0E}}{r}\right)^l P_{lk}\left(\cos\theta\right)
  \left[\Delta C_{lk}\cos\left(k\varphi\right)+\Delta S_{lk}\sin\left(k\varphi\right)\right],
\end{equation}
where $U_{b}$ is the Newtonian gravitational potential of body $b$, $\boldsymbol{{r}}_{bE}$ is the vector connecting the center of mass of body $b$ to that of the Earth, $\nabla U$ is the gradient of potential, $\Delta C_{lk}$ and $\Delta S_{lk}$ are the changes of normalized potential coefficients due to Earth tides. From Eqs.(\ref{grti1}) and (\ref{grti2}), the magnitude of the external-mass tidal potentials increases with satellite orbital altitude, while that of internal-mass tidal potential decreases correspondingly. The influence of tidal potentials can be characterized using Love numbers and more detailed computations can be found in Ref.\cite{petit2010iers,zhang2026influence,qin2020tidal}.

For the uplink, the tidal contribution to the frequency comparison is
\begin{equation}
y^{\mathrm{tid}}_{\mathrm{up}}
=
\frac{u_{\mathrm{tid}}(\boldsymbol{x}_2)-u_{\mathrm{tid}}(\boldsymbol{x}_1)}{c^2}.
\end{equation}
From an order-of-magnitude estimate, the tidal frequency shift in uplink can reach the level of $10^{-17}$ for low-orbit satellites (LEO), and for medium/geosynchronous Earth orbit (MEO/GEO) satellites, it may may approach the $10^{-16}$ level. Similarly, the downlink tidal contribution has approximately the opposite sign.

In the dual one-way DCS observable, the total tidal frequency shift is expressed as
\begin{equation}\label{grti3}
  y^{\text{tid}}_{\text{dcs}}=y^{\text{tid}}_{\text{do}}-y^{\text{tid}}_{\text{up}}\simeq 2\left(\frac{u_{\text{tid}}(\boldsymbol{{x}}_4)}{c^2}-\frac{u_{\text{tid}}(\boldsymbol{{x}}_3)}{c^2}\right),
\end{equation}
where the approximation holds when the uplink and downlink are nearly synchronized. The total tidal frequency shift in the dual one-way DCS is twice that of a one-way link. From Eqs.(\ref{grti1}) and (\ref{grti2}), external-mass tidal contribution dominate for case of MEO/GEO satellites.

\subsection{Clock synchronization and onboard time delay}
In the dual one-way DCS, the frequency shift data of the uplink is recorded onboard the satellite, while downlink data is recorded at the ground station. Gravitational redshift tests are obtained by combining these two data sets in the dual one-way DCS. So, any time-synchronization error between the ground and satellite clocks, denoted by $\delta t$, leads to a mismatch in the combination of uplink and downlink measurements. Clock synchronization refers to the determination of the time offset between the satellite and ground clocks, or equivalently the relation between their local time tags and a common coordinate-time scale, at the epochs associated with the two one-way measurements. To first order, the synchronization-induced error can be obtained by a Taylor expansion of the Doppler frequency shift with respect to the time-tag offset $\delta t$. Referenced to ground station time $t_4$, this error can be written as
\begin{equation}\label{grcs1}
  y^{\text{cs}}_{\text{dcs}}=-\frac{\boldsymbol{{n}}_{12} \cdot \boldsymbol{{a}}_{sg}(t_4)}{c}\delta t-\frac{\boldsymbol{{v}}^2_{12}(t_4)}{cD_{12}}\delta t+\frac{(\boldsymbol{{n}}_{12} \cdot\boldsymbol{{v}}_{12}(t_4))^2}{cD_{12}}\delta t.
\end{equation}
This contribution propagates directly into the DCS observable and therefore sets a requirement on the synchronization accuracy. This equation quantify clock synchronization requirements for the dual one-way DCS.
For example, in the ACES/CSS-like configurations, the clock comparison with the accuracy of $10^{-16}$ typically requires clock synchronization with the level 10 ns, and the clock comparison with the accuracy of $10^{-18}$ requires clock synchronization with the level $0.1$ ns. Recent advances in time and frequency transfer suggest that synchronization at the femtosecond level may become achievable in the future \cite{chen2024dual,caldwell2023quantum,deschenes2016synchronization,bergeron2019femtosecond,delva2012time,PhysRevA.99.023844}.

The difference between the satellite and ground clock readings is not assumed to remain constant. Over an observation interval, it may be represented locally as
\begin{equation}\label{cd11}
\Delta T(t) = \Delta T_0 +y_0(t-t_0)
+\frac{1}{2}\dot{y}(t-t_0)^2
+\delta T_{n}(t),
\end{equation}
where $\Delta T_0$ is the initial time offset at the reference time $t_0$, $y_0$ is the relative fractional-frequency offset, $\dot{y}$ is the frequency-drift rate, and $\delta T_{n}$ the stochastic timing error arising from clock and time-transfer noise. For a specific mission, these parameters can be estimated from the corresponding clock-comparison model and data.
Modern atomic and optical clocks may exhibit small residual deterministic drift after atomic referencing and calibration.
The high stability of the frequency standards alone does not imply that the inter-clock time offset is constant. The quadratic drift term may be neglected over a particular observation
interval if the corresponding contribution remains below the adopted uncertainty threshold. For longer observation intervals, the relative frequency offset and drift should be retained or estimated jointly with the physical parameters of interest.

In addition, the onboard time delay $\Delta t=t_3-t_2$ introduces an extra bias in the dual one-way DCS observable. To leading order, the corresponding contribution can be approximated as
\begin{equation}\label{grcs2}
  y^{\text{otd}}_{\text{dcs}}=-\frac{\boldsymbol{{n}}_{34}\cdot\boldsymbol{{a}}_{sg}(t_4)}{c}\Delta t.
\end{equation}
This contribution is dependent on the relative acceleration of the ground station and satellite. Under ideal conditions ($\Delta t=0$), this term vanishes. Typically, the uncertainty in the onboard delay is often smaller than the synchronization error. For ACES/CSS-like missions, an uncertainty of $0.1\ \mathrm{ns}$ corresponds to a frequency-shift contribution at approximately the $10^{-18}$ level. As a rough estimate, synchronization errors at the $10\ \mathrm{ns}$ and $0.1\ \mathrm{ns}$ levels would correspond to fractional uncertainties of about $10^{-6}$ and $10^{-8}$, respectively, in a gravitational-redshift test.

\subsection{Clock noise}

Clock noise is an important limiting factor in satellite-ground clock comparison. The output of an atomic clock can be modeled as a superposition of statistically independent noise processes. Characterizing the stochastic noise properties, the one-sided power spectral density of fractional frequency fluctuations can be expressed as \cite{lesage1979characterization}
\begin{equation}\label{gror1}
S(f)=\sum_{n=-2}^{2} h_n f^n,
\end{equation}
where $f$ is the Fourier frequency variable and $h_n$ is the noise coefficient associated with the corresponding power-law component. The five fundamental noise types are white phase modulation noise ($f^{2}$), flicker phase modulation noise ($f^{1}$), white frequency modulation noise ($f^{0}$), flicker frequency modulation noise ($f^{-1}$), and random-walk frequency modulation noise ($f^{-2}$). Based on the five types of clock noises, we can model and analyse the atomic clock frequency in the precision metrology application.

\section{Application in the Gravitational Redshift Tests}\label{sec4}

In this section, we perform numerical simulations of the dual one-way DCS observable for representative satellite-ground clock-comparison configurations. The observable is formed by combining one uplink and one downlink frequency-transfer measurement. In the dual one-way DCS observable, the gravitational-redshift contribution appears with an approximate factor of two. This feature is useful for extracting the gravitational-redshift signal, but it does not by itself imply a factor-of-two improvement in the final experimental sensitivity, which is also limited by residual Doppler terms, propagation effects, clock noise, orbit-determination errors, and clock-synchronization errors.

At present, the accuracy and stability of onboard atomic clocks in space missions are approaching the $10^{-16}$ level. Meanwhile, state-of-the-art ground-based optical atomic clocks have demonstrated accuracy and stability at the $10^{-18}$ level, and space optical clocks are also being developed toward this target. In the following, we apply the dual one-way DCS to two representative configurations: a CSS/ACES-like low-Earth-orbit configuration and a geosynchronous Earth orbit configuration. We estimate the achievable gravitational redshift tests for clock-comparison accuracies of $1\times10^{-16}$ and $1\times10^{-18}$. Although the relevant technologies have advanced substantially, an integrated satellite-ground optical-clock comparison at the $10^{-18}$ level remains a future capability. The configurations considered here are regarded as representative mission scenarios used to establish relativistic modeling and error-control requirements.

For realistic implementations, we assume that the satellite position and velocity can be determined with uncertainties of 0.1 m and 0.1 mm/s, respectively. The position and velocity of the ground station are assumed to be known with better accuracy, so that the dominant orbit-related uncertainties are satellite dominated. Considering the optical communication channel, we set the observation cutoff elevation angle for the satellites on the ground station to 10$^{\circ}$ and assume that frequency observables and satellite orbit determination data are available during the observation period. The numerical values listed below should be interpreted as the order-of-magnitude estimates under these assumptions.

\begin{figure}
\includegraphics[width=0.7\textwidth]{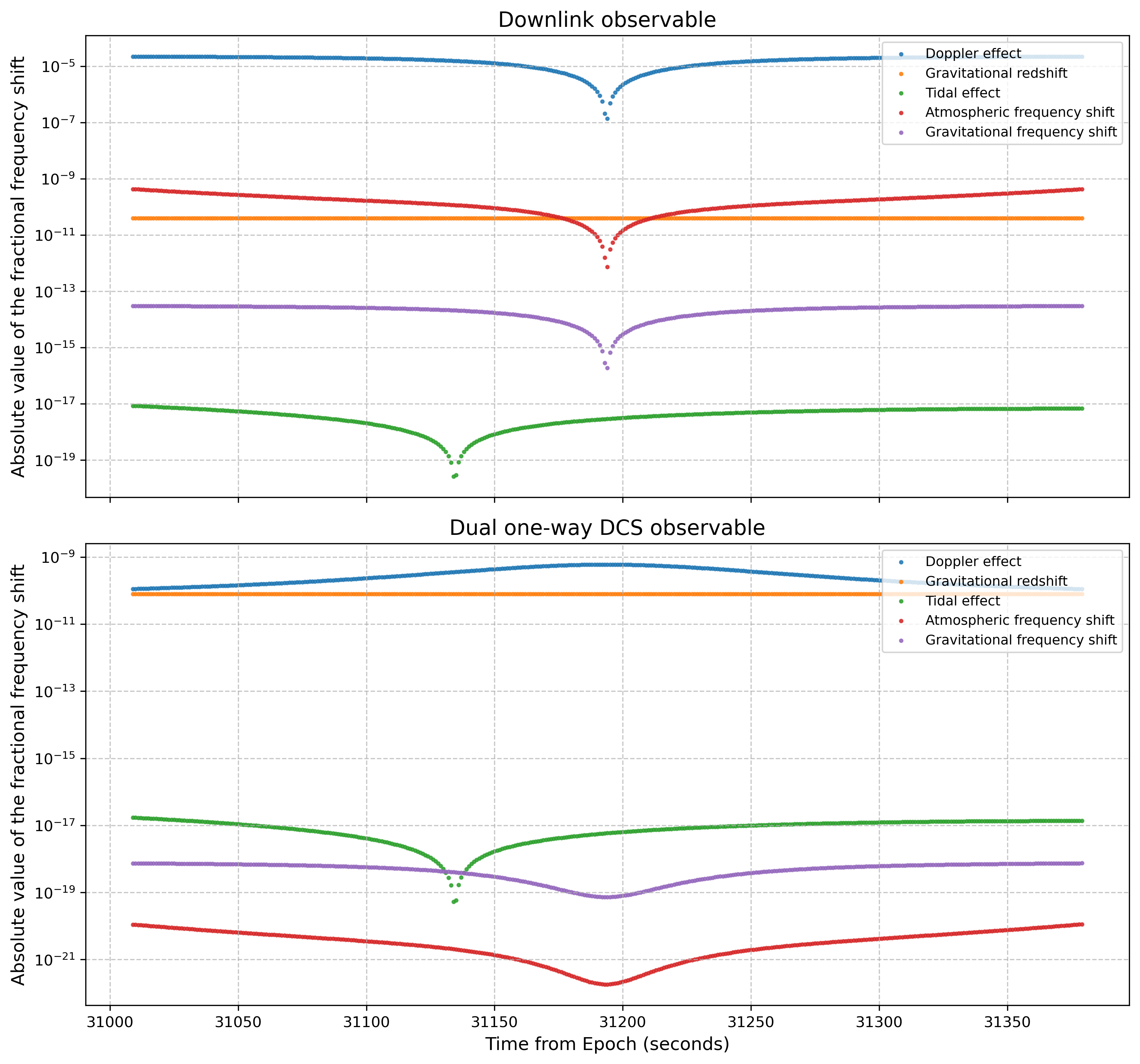}
\caption{\label{fig2} The absolute value of the fractional frequency shift for the main effects in the CSS configurations. The downlink observable is plotted in the upper figure and dual one-way DCS observalbe is plotted in the lower figure.}
\end{figure}

\begin{table}
\caption{\label{tab2}. Order-of-magnitude estimations of the main effects in the CSS/ACES experiment. The first column represents the types of effects in clock comparison. The second column gives the magnitude in the one-way observable, the third column gives the magnitude in the dual DCS observable, and the fourth column gives the estimated residual uncertainty after modeling.}
\begin{ruledtabular}
\begin{tabular}{cccc}
{Terms} &One-way observable    &Dual one-way DCS observable  &Error\\
\hline
{Doppler effect} &$\sim2\times10^{-5}$   &$\sim6\times10^{-10}$   &$\sim 1\times10^{-17}$ \\
{Gravitational redshift signal}  &$\sim4\times10^{-11}$   &$\sim8\times10^{-11}$   &$\sim1\times10^{-17}$ \\
{Atmospheric frequency shift}  &$\sim4\times10^{-10}$   &$<1\times10^{-18}$   &$<1\times10^{-18}$ \\
{Shapiro gravitational frequency shift}  &$\sim3\times10^{-14}$   &$<1\times10^{-18}$   &$<1\times10^{-18}$ \\
{Tidal effects}  &$\sim8\times10^{-18}$   &$\sim2\times10^{-17}$   &$<1\times10^{-18}$ \\
\end{tabular}
\end{ruledtabular}
\end{table}

In the CSS/ACES configurations, the altitude of the space station is about 400 km, and the ground station is taken to be the Beijing station. For this configuration, the gravitational redshift signal in the dual one-way DCS observable is approximately $8\times10^{-11}$. Based on the models discussed in Section.\ref{sec3}, we simulate the main contributions to the downlink observable (upper) and dual one-way DCS observable (lower), as shown in FIG.\ref{fig2}. The blue solid dot represents doppler effect, orange solid dot represents the gravitational redshift signal, green solid dot represents the tidal contribution, red solid dot represents the atmospheric contribution and purple solid dot represents Shapiro gravitational delay contribution. The corresponding magnitudes and estimated uncertainties are summarized in TABLE.\ref{tab2}.

For the CSS/ACES case, the first-order Doppler effect is suppressed by the dual one-way DCS, leaving a residual Doppler contribution at approximately the $6\times10^{-10}$ level. With the assumed satellite orbit-determination accuracy, the uncertainty of the residual Doppler correction is estimated to be $2\times10^{-17}$. The systematic uncertainties associated with atmospheric delay, Shapiro delay, and tidal effects are below the $10^{-16}$ level and can be reduced to below the $10^{-18}$ level after modeling.
For a clock-comparison accuracy of $1\times10^{-16}$, the expected uncertainty of the gravitational redshift test is at $1\times10^{-6}$ level.
For a clock-comparison accuracy approaching $1\times10^{-18}$, the clock noise itself is no longer the only limiting factor. The uncertainties of residual Doppler and gravitational potential become important. If the satellite velocity determination is further improved so that the residual Doppler uncertainty can be suppressed to the $10^{-18}$ level, while the other systematic uncertainties remain below $10^{-17}$, a gravitational redshift test at the $10^{-7}$ level becomes feasible for the CSS/ACES configuration. In this case, the limiting factor is expected to be the modeling accuracy of the Earth's gravitational potential.

\begin{figure}
\includegraphics[width=0.7\textwidth]{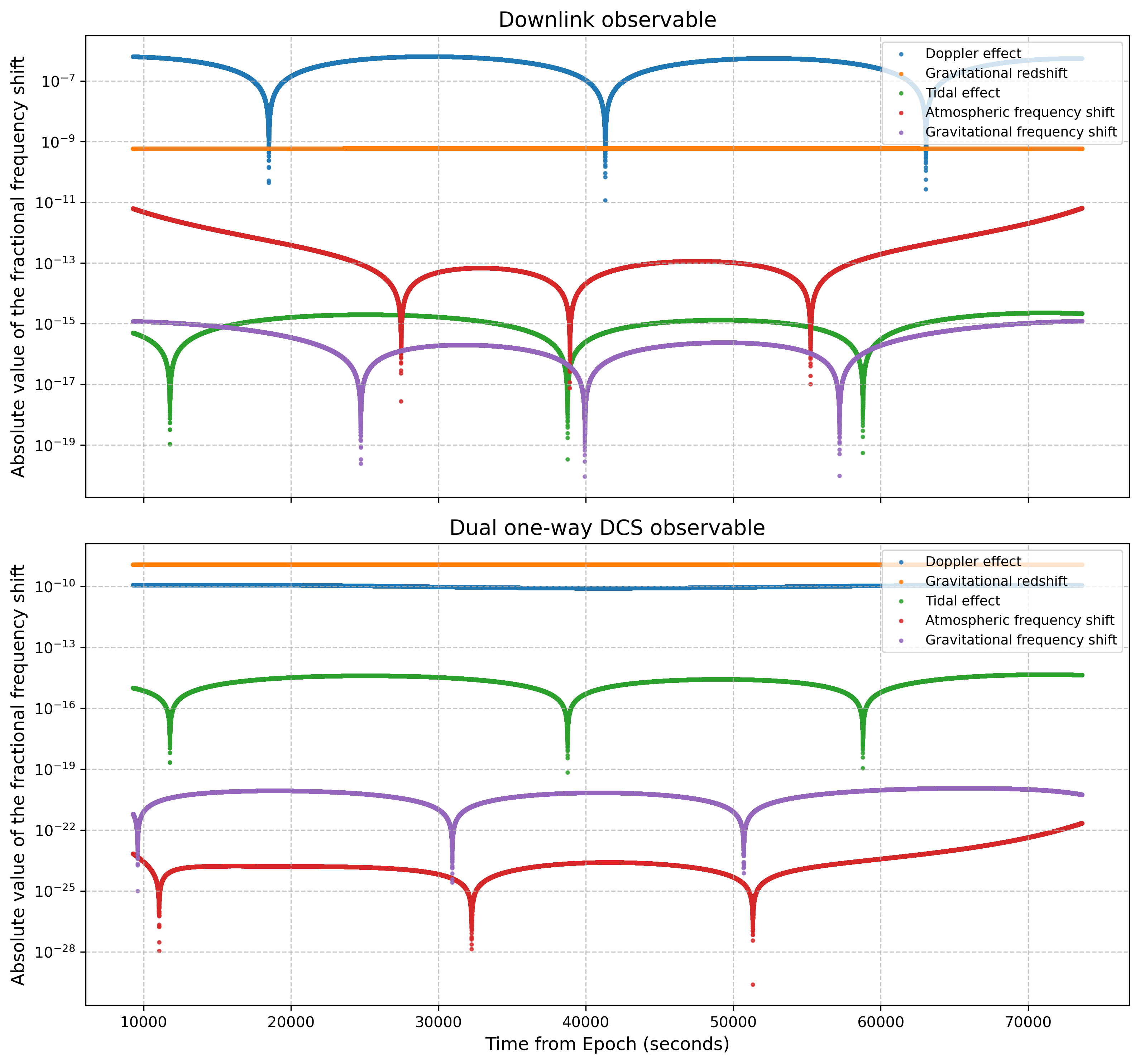}
\caption{\label{fig3} The absolute value of the fractional frequency shift for the main effects in the GEO configurations. The downlink observable is plotted in the upper figure and dual one-way DCS observalbe is plotted in the lower figure.}
\end{figure}

\begin{table}
\caption{\label{tab3} Order-of-magnitude estimates of the main effects in the GEO experiment.  }
\begin{ruledtabular}
\begin{tabular}{cccc}
{Terms} &One-way observable    &Dual one-way DCS observable  &Error\\
\hline
{Doppler effect} &$\sim6\times10^{-7}$   &$\sim1\times10^{-10}$   &$\sim7\times10^{-18}$ \\
{Gravitational redshift signal}  &$\sim6\times10^{-10}$   &$\sim1\times10^{-9}$   &$\sim1\times10^{-17}$ \\
{Atmospheric frequency shift}  &$\sim6\times10^{-12}$   &$<1\times10^{-18}$   &$<1\times10^{-18}$ \\
{Shapiro gravitational frequency shift}  &$\sim1\times10^{-15}$   &$<1\times10^{-18}$   &$<1\times10^{-18}$ \\
{Tidal effects}  &$\sim2\times10^{-15}$   &$\sim4\times10^{-15}$   &$<1\times10^{-18}$ \\
\end{tabular}
\end{ruledtabular}
\end{table}

As a second representative orbit, we consider a geostationary-Earth-orbit configuration. This example provides a larger gravitational-potential difference and a dynamical regime complementary to the low-Earth-orbit case. In this GEO configuration, the altitude of the satellite is about 36,000 km, the inclination is about 53$^\circ$ , and the ground station is again taken to be the Beijing station. Owing to the larger gravitational-potential difference between the satellite and the ground station, the gravitational redshift signal in the dual one-way DCS observable increases to approximately $1.2\times10^{-9}$. The simulated downlink and dual oway DCS observables are shown in FIG.\ref{fig3},  and the corresponding magnitudes and uncertainties are summarized in TABLE.\ref{tab3}.

In the GEO configuration, the residual Doppler contribution in the dual one-way DCS is approximately $1\times10^{-10}$ and the corresponding uncertainty can be reduced to $7\times10^{-18}$ level under the assumed orbit-determination accuracy. The atmospheric and Shapiro-gravitational-delay residuals remain at or below the $10^{-18}$ level after cancellation. The tidal contribution is enhanced because of the larger orbital altitude. Its modeling uncertainty can remain below the $10^{-18}$ level with appropriate tidal models.
For a clock-comparison accuracy of $1\times10^{-16}$, the expected gravitational redshift test reaches $1\times10^{-7}$ level. This represents a improvement over the low-Earth-orbit configuration, mainly because of the larger gravitational redshift signal at GEO altitude.
For a clock-comparison accuracy of $1\times10^{-18}$, the expected gravitational redshift test is at $1\times10^{-8}$ level under the assumption that the dominant systematic uncertainties can be controlled at or below the $10^{-17}$ level. In this regime, the limiting factor is expected to be the accuracy of the Earth's gravitational field model rather than the clock noise itself.

For satellite-ground clock comparisons targeting the $10^{-18}$ level, the improving of satellite orbit determination and velocity measurement are essential for suppressing the uncertainty of the residual Doppler correction below $10^{-18}$. At the same time, the Earth's gravitational potential must be modeled with sufficient accuracy. Since an uncertainty of $0.1$ m$^2$/s$^2$ in the gravitational potential corresponds to a fractional frequency uncertainty of about $1\times10^{-18}$, the geopotential model becomes a key limiting factor for future high-precision redshift tests. Conversely, high-accuracy satellite-ground clock-comparison data may also provide a way to improve the Earth's gravitational field model and calibrate the corresponding geopotential coefficients.



%

\section{conclusion}\label{sec5}

In this work, we have investigated a dual one-way Doppler-cancellation scheme (DCS) for gravitational-redshift tests in satellite-ground clock comparisons. We have developed the corresponding relativistic observable up to order $c^{-3}$, making the model suitable for future satellite-ground experiments employing optical clocks at the $10^{-18}$ level. In this configuration, the observable is formed from two one-way signal links, namely an uplink and a downlink, with the corresponding frequency data recorded independently at the satellite and at the ground station. The dual one-way DCS observable is then constructed by combining these two measurements in post-processing. Based on this framework, we have analyzed the main contributions to the observable, including the first- and higher-order Doppler terms, the gravitational-redshift term, neutral-atmospheric and ionospheric effects, Shapiro delay, tidal effects, clock-synchronization errors, and clock noise. Our results show that the dual one-way DCS suppresses the dominant first-order Doppler contribution while retaining the gravitational-redshift signal, and provides a compact two-link alternative to conventional three-link Doppler-cancellation configurations. At the same time, this simplification may place stringent requirements on synchronization between the satellite and ground clocks.

To demonstrate the performance of the formulation, we simulate the gravitational redshift tests in the LEO (ACES/CSS) and GEO satellites. In the DCS observables, the gravitational redshift signals can reach $8\times10^{-11}$ and $1\times10^{-9}$ for the LEO and GEO satellites. These results indicate that gravitational redshift tests at the $10^{-6}$ to $10^{-8}$ level are achievable, depending on the clock performance and orbit-determination accuracy. Our analysis further shows that the influences of residual Doppler effect, atmospheric delay, Shapiro delay, and tidal effects can all be controlled at or below the $10^{-17}$-$10^{-18}$ with current or near-future technologies. The key requirement is sub-nanosecond clock synchronization. In particular, a synchronization accuracy of approximately $0.1$ ns is required to support $10^{-18}$-level satellite-ground clock comparisons when combined with appropriate orbit determination and time-transfer techniques.

The dual one-way DCS therefore provides a useful relativistic framework and error-modeling reference for next-generation space-based gravitational-redshift tests at the $10^{-8}$ level or beyond. The redshift signal may also provide sensitivity to possible deviations from metric gravity, fifth-force effects, and  dark-matter-induced signatures. Realizing this potential will require further improvements in Earth's gravitational model, orbit determination, optical time transfer, and clock synchronization. Overall, the dual one-way DCS offers a promising and experimentally practical architecture for future space-based tests of fundamental physics.

\section{Acknowledgment}
This work is supported by the National Natural Science Foundation of China (Grants No.12305062), Strategic Priority Research Program on Space Science, the Chinese Academy of Sciences (XDA30040400), and Fundamental Research Funds for the Central Universities, Sun Yat-sen University.

\begin{appendix}
\section{Doppler frequency shift in the dual one-way DCS }\label{app1}

This Appendix presents several relationships and results for the Doppler frequency shift in the dual one-way DCS. The one-way frequency transfer can be characterized as $f_{A}/f_{B}=d\tau_{B}/d\tau_{A}$, and the standard GR prediction have been studied some researches \cite{blanchet2001relativistic,PhysRevD.66.024045}. Ignoring the influences of atmospheric delay and Shapiro delay, the one-way frequency transfers of uplink and downlink can be expanded to the order $c^{-3}$
\begin{equation}\label{ow1}
\begin{aligned}
y_{\text{up}}=\frac{f_2}{f_1}
&= 1 - \frac{\boldsymbol{N}_{12} \cdot \boldsymbol{v}_{12}}{c}
   + \frac{1}{c^2} \Big[- \big( \boldsymbol{N}_{12} \cdot \boldsymbol{v}_{12} \big) \big( \boldsymbol{N}_{12} \cdot \boldsymbol{v}_{g}(t_1) \big)+\frac{\boldsymbol{v}^2_{s}(t_2)}{2}-\frac{\boldsymbol{v}^2_{g}(t_1)}{2} +w_{2}-w_{1}\Big] \\
&\quad - \frac{\big( \boldsymbol{N}_{12} \cdot \boldsymbol{v}_{12} \big)}{c^3} \Bigg[ \big( \boldsymbol{N}_{12} \cdot \boldsymbol{v}_{g}(t_1) \big)^2 + \frac{\boldsymbol{v}_{s}^2(t_2)}{2} - \frac{\boldsymbol{v}_{g}^2(t_1)}{2}  +w_{2}-w_{1}\Bigg] +O(c^{-4}),
\end{aligned}
\end{equation}
and
\begin{equation}\label{ow2}
\begin{aligned}
y_{\text{do}}=\frac{f_4}{f_3}
&= 1 - \frac{\boldsymbol{N}_{34} \cdot \boldsymbol{v}_{34}}{c}
   + \frac{1}{c^2} \Big[ -\big( \boldsymbol{N}_{34} \cdot \boldsymbol{v}_{34} \big) \big( \boldsymbol{N}_{34} \cdot \boldsymbol{v}_{s}(t_3) \big)
   +\frac{\boldsymbol{v}^2_{g}(t_4)}{2}-\frac{\boldsymbol{v}^2_{s}(t_3)}{2} +w_{4}-w_{3}\Big] \\
&\quad - \frac{\big( \boldsymbol{N}_{34} \cdot \boldsymbol{v}_{34} \big)}{c^3} \Bigg[ \big( \boldsymbol{N}_{34} \cdot \boldsymbol{v}_{s}(t_3)  \big)^2 + \frac{\boldsymbol{v}_{g}^2(t_4)}{2} - \frac{\boldsymbol{v}_{s}^2(t_3)}{2}+w_{4}-w_{3} \Bigg]+O(c^{-4}),
  \end{aligned}
\end{equation}
where $w$ represent the scalar potential that is composed of the Earth gravitational potential $U_{E}$ and tidal potential $u_{\text{tid}}$, $\boldsymbol{{N}}=\boldsymbol{{R}}/R$ is the vector of coordinate distance $\boldsymbol{{R}}$, and $\boldsymbol{{v}}_{lm}$ is the relative velocity.

Considering two one-way frequency transfers in the uplink and downlink, the dual one-way DCS observable is given by $y_{\text{dcs}}=y_{\text{do}}-y_{\text{up}}$. Generally, all quantities in $y_{\text{dcs}}$ should be expressed in terms of parameters referenced to the reception time $t_{4}$.
It is necessary to expressed the unit vector $\boldsymbol{N}_{12}$ in terms of the unit vector $\boldsymbol{N}_{34}$. The relationship between the coordinate distances of uplink and downlink is given by
\begin{equation}\label{ow3}
  \boldsymbol{R}_{12}=-\boldsymbol{R}_{34}+\boldsymbol{v}_{g}\left(t_{4}\right)\cdot T_{14}-\frac{1}{2}\boldsymbol{a}_{g}\left(t_{4}\right)\cdot T_{14}^{2}-\Delta\boldsymbol{r}_{s}\left(\Delta t\right),
\end{equation}
where $\Delta\boldsymbol{r}_{s}\left(\Delta t\right)=\boldsymbol{v}_{s}(t_2) \Delta t+(1/2)\boldsymbol{a}_{s}(t_2) \Delta t^2$ is the satellite displacement in the onboard time delay $\Delta t$.
From this equation, we obtain
\begin{equation}\label{appe1}
\begin{aligned}
 \boldsymbol{N}_{12} &= -\boldsymbol{N}_{34}\left[1+2\frac{\boldsymbol{N}_{34}\cdot\boldsymbol{v}_g(t_4)}{c}
 +\frac{\boldsymbol{N}_{34}\cdot\boldsymbol{v}_{34}}{R_{34}}\Delta t -\frac{1}{c^{2}}\left[4\left(\boldsymbol{N}_{34}\cdot \boldsymbol{v}_{g}(t_{4})\right)^{2}-2\boldsymbol{v}_{g}^{2}(t_{4})-\boldsymbol{a}_{g}(t_{4})\cdot \boldsymbol{R}_{34}\right]\right]\\
 & +\frac{\boldsymbol{v}_g\left(t_4\right)}{c}\left(2+2\frac{\boldsymbol{N}_{34}\cdot\boldsymbol{v}_g\left(t_4\right)}{c}
 +\frac{\boldsymbol{N}_{34}\cdot\boldsymbol{v}_{34}}{R_{34}}\Delta t\right)
  -2\frac{\boldsymbol{a}_{g}\left(t_{4}\right)R_{34}}{c^{2}}+\left(\frac{\boldsymbol{v}_{34}}{R_{12}}
  +\frac{\boldsymbol{a}_{s}\left(t_{4}\right)}{c}-\frac{2\boldsymbol{a}_{g}\left(t_{4}\right)}{c}\right)\Delta t.
\end{aligned}
\end{equation}

We also need reexpress the velocity of the satellite and ground station at emission in terms of the quantities at reception time, as follows
\begin{equation}\label{appe2}
  \boldsymbol{v}_{s}\left(t_{2}\right)=\boldsymbol{v}_{s}\left(t_{3}\right)-\boldsymbol{a}_{s}\left(t_{3}\right)\Delta t+\frac{1}{2}\boldsymbol{b}_{s}\left(t_{3}\right)\Delta t^{2}
\end{equation}

\begin{equation}\label{appe3}
\boldsymbol{v}_g\left(t_1\right)
=\boldsymbol{v}_g(t_4)-\frac{2R_{34}}{c}\boldsymbol{a}_g(t_4)+\frac{2}{c^2}\left[\boldsymbol{b}_g(t_4)R_{34}^2
+\left(\boldsymbol{N}_{34}\cdot\boldsymbol{v}_g(t_4)\right)\boldsymbol{a}_g(t_4)R_{34}\right]-\boldsymbol{a}_g(t_4)\Delta t+\frac{\boldsymbol{a}_g(t_4)}{c}\left(\boldsymbol{N}_{34}\cdot\boldsymbol{v}_{34}\right)\Delta t.
\end{equation}

In addition, introducing the instantaneous distance $\boldsymbol{{D}}_{34}=\boldsymbol{{x}}_{g}(t_{4})-\boldsymbol{{x}}_{s}(t_{4})$ with its unit vector $\boldsymbol{{n}}_{34}$, one can express the term $\boldsymbol{{N}}_{34}$ in terms of the $\boldsymbol{{n}}_{34}$ as
\begin{equation}\label{app111}
  \boldsymbol{{N}}_{34}=\boldsymbol{{n}}_{34}
  \left(1-\frac{\boldsymbol{{n}}_{34}\cdot\boldsymbol{{v}}_{s}(t_3)}{c}\right)+\frac{\boldsymbol{{v}}_{s}(t_3)}{c}.
\end{equation}

Taking these relativistic expansions into Eq.(\ref{gr6}), the dual one-way DCS can be reexpressed in terms of the quantities at reception time $t_4$. After a long computation, we obtain
\begin{equation}\label{appe4}
  y_{\text{dcs}}=y^{\text{gr}}_{\text{dcs}}+y^{\text{tid}}_{\text{dcs}}+y^{\text{dop2}}_{\text{dcs}}+y^{\text{dop3}}_{\text{dcs}}+
  y^{\text{otd}}_{\text{dcs}},
\end{equation}
where the first term is the gravitational redshift, the second term represents the tidal influence, the third term is the $c^{-2}$-order residual Doppler frequency shift, the fourth term is the $c^{-3}$-order residual Doppler frequency shift, and fifth term indicates the influences of the onboard time delay $\Delta t = t_3- t_2$, which is the coordinate-time separation between the reception of the uplink and the emission of the downlink at the satellite.
The $y^{\text{gr}}_{\text{dcs}}$ is given by
\begin{equation}\label{appgr}
  y^{\text{gr}}_{\text{dcs}}=\frac{1}{c^2} \left[U_{E}(\boldsymbol{x}_{4})+U_{E}(\boldsymbol{x}_{1})-U_{E}(\boldsymbol{x}_{2})-U_{E}(\boldsymbol{x}_{3})\right].
\end{equation}
The term $y^{\text{tid}}_{\text{dcs}}$ is given by
\begin{equation}\label{apptid}
  y^{\text{tid}}_{\text{dcs}}= \frac{1}{c^2} \left[u_{\text{tid}}(\boldsymbol{{x}}_4)+u_{\text{tid}}(\boldsymbol{{x}}_1)
  -u_{\text{tid}}(\boldsymbol{{x}}_3)-u_{\text{tid}}(\boldsymbol{{x}}_2)\right].
\end{equation}
The term $y^{\text{dop2}}_{\text{dcs}}$ is given by
\begin{equation}\label{apped2}
  y^{\text{2}}_{\text{dcs}}=-\frac{2\boldsymbol{{D}}_{34}\cdot\boldsymbol{{a}}_{g}(t_4)}{c^2}
+\frac{(\boldsymbol{{n}}_{34}\cdot\boldsymbol{{v}}_{34}(t_4))^{2}}{c^2}
-2\frac{(\boldsymbol{{v}}_{g}(t_4)\cdot\boldsymbol{{v}}_{34}(t_4))}{c^2},
\end{equation}
where $\boldsymbol{{v}}_{34}(t_4)=\boldsymbol{{v}}_{g}(t_4)-\boldsymbol{{v}}_{s}(t_4)$ is the relative velocity between satellite and ground station at time $t_4$.
The term $y^{\text{dop3}}_{\text{dcs}}$ is given by
\begin{equation}\label{apped3}
\begin{aligned}
 y^{\text{dop3}}_{\text{dcs}}& =2\frac{\boldsymbol{D}_{34}\cdot\boldsymbol{b}_{g}(t_4)}{c^3}D_{34}
 +4\frac{\boldsymbol{a}_{g}(t_4)\cdot\boldsymbol{v}_{g}(t_4)D_{34}}{c^3}
 -2\frac{\boldsymbol{a}_{s}\left(t_4\right)\cdot\boldsymbol{v}_{34}(t_4)D_{34}}{c^3}
  +\frac{\left(\boldsymbol{n}_{34}\cdot\boldsymbol{v}_{34}(t_4)\right)}{c^3}\times \\
 & \biggl[3\Big(\boldsymbol{n}_{34}\cdot\boldsymbol{v}_{{g}}(t_4)\Big)^2+3\Big(\boldsymbol{n}_{34}\cdot\boldsymbol{v}_{s}\left(t_4\right)\Big)^2
 -\boldsymbol{v}_{34}^2(t_4)-2\boldsymbol{v}_{s}^2(t_4)+\boldsymbol{a}_{g}(t_4)\cdot\boldsymbol{D}_{34}
 +2\boldsymbol{a}_{s}(t_4)\cdot\boldsymbol{D}_{34}\biggr]  \\
&+ \frac{\boldsymbol{{n}}_{34}\cdot\boldsymbol{{v}}_{34}(t_4)}{c^3}\left[  \frac{\boldsymbol{v}_{s}^2(t_2)}{2} + \frac{\boldsymbol{v}_{s}^2(t_3)}{2}- \frac{\boldsymbol{v}_{g}^2(t_1)}{2} - \frac{\boldsymbol{v}_{g}^2(t_4)}{2}+w_{2}+w_{3} -w_{1}-w_{4} \right].
\end{aligned}
\end{equation}
The term $y^{\text{otd}}_{\text{dcs}}$ is given by
\begin{equation}\label{appe6}
\begin{gathered}
y^{\text{otd}}_{\text{dcs}}=-\frac{\boldsymbol{n}_{34}\cdot\boldsymbol{a}_{sg}(t_4)}{c}\Delta t+\frac{\boldsymbol{n}_{34}\cdot\boldsymbol{v}_{34}\left(t_4\right)}{c}\left[\left(\frac{\boldsymbol{n}_{34}\cdot\boldsymbol{v}_{34}(t_4)}{R_{34}}\Delta t+\frac{\boldsymbol{n}_{34}\cdot\boldsymbol{a}_s(t_4)}{c}\Delta t+\frac{\boldsymbol{n}_{34}\cdot\boldsymbol{a}_g(t_4)}{c}\Delta t\right)\right] \\
-\frac{\boldsymbol{v}_g\left(t_4\right)\cdot\boldsymbol{v}_{34}\left(t_4\right)}{c^2}\frac{\boldsymbol{n}_{34}\cdot\boldsymbol{v}_{34}\left(t_4\right)}{R_{34}}\Delta t-\frac{\boldsymbol{v}_{34}^2\left(t_4\right)}{cR_{12}}\Delta t-\frac{\boldsymbol{a}_s\left(t_4\right)\cdot\boldsymbol{v}_{34}\left(t_4\right)}{c^2}\Delta t+\frac{2\boldsymbol{a}_g\left(t_4\right)\cdot\boldsymbol{v}_{34}\left(t_4\right)}{c^2}\Delta t \\
-\frac{2\left(\boldsymbol{n}_{34}\cdot\boldsymbol{v}_{34}(t_4)\right)\left(\boldsymbol{n}_{34}\cdot\boldsymbol{a}_\mathrm{g}(t_4)\right)}{c^2}\Delta t-2\frac{\left(\boldsymbol{n}_{34}\cdot\boldsymbol{v}_\mathrm{g}(t_4)\right)\left(\boldsymbol{n}_{34}\cdot\boldsymbol{a}_\mathrm{sg}(t_4)\right)}{c^2}\Delta t+\frac{2\boldsymbol{v}_\mathrm{g}(t_4)\cdot\boldsymbol{a}_\mathrm{sg}(t_4)}{c^2}\Delta t.
\end{gathered}
\end{equation}

\end{appendix}


%

\end{document}